\documentstyle[multicol,aps,psfig]{revtex}

\begin{document}

\title{Heavy Quark Energy Loss in Nuclear Medium}
\author{Ben-Wei Zhang$^{a}$, Enke Wang$^{a}$ and Xin-Nian Wang$^{b,c}$}
\address{$^a$ Institute of Particle Physics, Huazhong Normal University,
         Wuhan 430079, China}
\address{$^b$Nuclear Science Division, MS 70R0319,
Lawrence Berkeley National Laboratory, Berkeley, CA 94720 USA}
\address{$^c$Department of Physics, Shandong University,
         Jinan 250100, China}

\date{September 5, 2003}

\maketitle

%\vspace{-2.0in}
%\vspace{2.in}

\begin{abstract}
\baselineskip=12pt Multiple scattering, modified fragmentation 
functions and radiative energy loss of a heavy quark propagating 
in a nuclear medium are investigated in perturbative QCD.
Because of the quark mass dependence of the gluon formation time, 
the medium size dependence of heavy quark energy loss is found
to change from a linear to a quadratic form when the initial energy
and momentum scale are increased relative to the quark mass. The
radiative energy loss is also significantly suppressed relative to
a light quark due to the suppression of collinear gluon emission by 
a heavy quark.
\end{abstract}


\begin{multicols}{2}

%\section{Introduction}

An energetic parton propagating in a dense medium suffers a large amount
of energy loss due to multiple scattering and induced gluon 
bremsstrahlung\cite{GW1}. In a static medium, the total energy loss of
a massless parton (light quark or gluon) is found to have a quadratic
dependence on the medium size \cite{BDMPS,Zh,GLV,Wie,GuoW} due to
non-Abelian Landau-Pomeranchuk-Migdal (LPM) interference effect.
In an expanding medium, the total energy loss can be cast into
a line integral weighted with local gluon density along the parton 
propagation path \cite{gvw,ww02,cw}. Therefore, the measurement 
of parton energy loss can be used to study 
properties of the medium similar to 
the technique of computed tomography. Recent experimental
measurements \cite{phenix,star} of centrality dependence of
high-$p_T$ hadron suppression agree very well \cite{wang03} 
with such a parton energy loss mechanism.

Because of the large mass of the heavy quark with a velocity 
$v\approx 1-M^2/2E^2$, the formation time of gluon radiation, 
$\tau_f \sim 1/(\omega_gM^2/2E^2+\ell_T^2/2\omega_g)$ is reduced 
relative to a light quark. One should then expect the LPM 
effect to be significantly reduced for intermediate 
energy heavy quarks. In addition, the heavy quark mass also 
suppresses gluon radiation amplitude at angles smaller
than the ratio of the quark mass to its energy \cite{kharzeev} 
relative to the gluon radiation off a light quark. Both mass 
effects will lead to a heavy quark energy loss different 
from a light quark propagating in a dense
medium. This might explain why one has not observed
significant heavy quark energy loss from the PHENIX \cite{phenix2}
measurement of the single electron spectrum from charm production
in $Au+Au$ collisions at $\sqrt{s}=130$ GeV. In this Letter, 
we report a study on medium induced energy loss and the
modified fragmentation function of a heavy quark. In particular,
we will show how the mass effects reduce the total energy loss and
how the medium size dependence changes from a linear dependence to
a quadratic one when the energy of the heavy quark or the momentum
scale is increased. Similar results have been reported in Ref. \cite{gyucharm}
during the completion of this work.

%\section{Generalized Factorization}

To separate the complication of heavy quark production and
propagation, we consider a simple process of charm quark
production via the charge-current interaction in DIS off a large
nucleus. The results can be easily extended to heavy quark 
propagation in other dense media.
The differential cross section for the semi-inclusive process
$\ell(L_1) + A(p) \longrightarrow \nu_{\ell}(L_2) + H(\ell_H) +X$ 
can be expressed as
\begin{equation}
E_{L_2}E_{\ell_H}\frac{d\sigma_{\rm DIS}}{d^3L_2d^3\ell_H}
=\frac{G_{\rm F}^2}{(4\pi)^3 s} L_{\mu\nu}^{cc}
E_{\ell_H}\frac{dW^{\mu\nu}}{d^3\ell_H} \; .
\label{sigma}
\end{equation}
Here $L_1$ and $L_2$ are the four momenta of the incoming lepton and 
the outgoing neutrino, $\ell_H$ the observed heavy quark meson momentum,
$p = [p^+,m_N^2/2p^+,{\bf 0}_\perp]$ is the momentum per nucleon in 
the nucleus, and $s=(p+L_1)^2$. $G_{\rm F}$ is the
four-fermion coupling constant
and $q =L_2-L_1 = [-Q^2/2q^-, q^-, {\bf 0}_\perp]$ the momentum transfer
via the exchange of a $W$-boson. The charge-current leptonic 
tensor is given by
$L_{\mu\nu}^{cc}=1/2{\rm Tr}({\not\!L_1}\gamma_{\mu}
(1-\gamma_5){\not\!L_2}(1-\gamma_5)\gamma_{\nu})$.
We assume $Q^2\ll M_W^2$.
The semi-inclusive hadronic tensor is defined as,
\begin{eqnarray}
E_{\ell_H}\frac{dW_{\mu\nu}}{d^3\ell_H}&=&
\frac{1}{2}\sum_X \langle A|J^+_\mu|X,H\rangle
\langle X,H| J_\nu^{+\dagger}|A\rangle \nonumber \\
&\times &2\pi \delta^4(q+p-p_X-\ell_H) \label{hadronic}
\end{eqnarray}
where $\sum_X$ runs over all possible final states and
$J^+_\mu=\bar{c} \gamma_\mu(1-\gamma_5)s_{\theta}$ is the
hadronic charged current. 
Here, $s_\theta=s\cos\theta_{\rm C}-d\sin\theta_{\rm C}$ and
$\theta_{\rm C}$ is the Cabibbo angle. To the leading-twist 
in collinear approximation, the semi-inclusive cross section 
factorizes into the product of quark distribution 
$f_{s_\theta}^A(x_B+x_M)$, the heavy quark fragmentation
function $D_{Q\rightarrow H}(z_H)$ ($z_H=\ell_H^-/\ell_Q^-$)
and the hard partonic part $H^{(0)}_{\mu\nu}(k,q,M)$ \cite{heavy2}.
Here, $x_B =Q^2/2p^+q^-$ is the Bjorken variable and $x_M=M^2/2p^+q^-$.

Similar to the case of light quark propagation in  nuclear 
medium \cite{GuoW}, the generalized factorization 
of multiple scattering processes \cite{LQS} will be employed.
We will only consider double parton scattering. The leading 
contributions are the twist-four terms that are enhanced by 
the nuclear medium in a collinear expansion, assuming a small 
expansion parameter $\alpha_s A^{1/3}/Q^2$. The evaluation of 23 
cut diagrams are similar to the case of a light quark \cite{heavy2}. 
The dominant contribution comes from the central cut diagram, giving
the semi-inclusive tensor for heavy quark fragmentation from double
quark-gluon scattering,
\begin{eqnarray}
\frac{W_{\mu\nu}^{D}}{dz_h} &=&\sum \,\int dx H^{(0)}_{\mu\nu}
\int_{z_h}^1\frac{dz}{z}D_{Q\rightarrow H}(\frac{z_H}{z})
\frac{C_A\alpha_s}{2\pi}  \frac{1+z^2}{1-z}\nonumber \\
&\times& \int \frac{d\ell_T^2}{[\ell_T^2+(1-z)^2 M^2]^4}\ell_T^4
\frac{2\pi\alpha_s}{N_c} T^{A,C}_{qg}(x,x_L,M^2)  \nonumber \\
&+& (g-{\rm frag.})+({\rm virtual\,\, corrections})\, ,
\label{wd1}
\end{eqnarray}
where
\begin{eqnarray}
 T^{A,C}_{qg}(&x,&x_L,M^2)= \frac{1}{2} \int
\frac{dy^{-}}{2\pi}\, dy_1^-dy_2^- \widetilde{H_C^D} \nonumber \\
&\times& \langle A | \bar{\psi}_q(0)\, \gamma^+\, F_{\sigma}^{\
+}(y_{2}^{-})\, F^{+\sigma}(y_1^{-})\,\psi_q(y^{-})
| A\rangle \nonumber \\
&\times & e^{i(x+x_L)p^+y^-}
 \theta(-y_2^-)\theta(y^- -y_1^-) \label{TC}
\end{eqnarray}
are twist-four quark-gluon correction functions of the nucleus.
Here
\begin{eqnarray}
\widetilde{H_{C}^D}&=& c_1(z,\ell_T^2,M^2)
(1-e^{-i\widetilde{x}_L p^+y_2^-})(1-e^{-i\widetilde{x}_L p^+(y^--y_1^-)})
 \nonumber  \\
&+& c_2(z,\ell_T^2, M^2)[e^{-i\widetilde{x}_L p^+y_2^-}
(1-e^{-i\widetilde{x}_L p^+(y^--y_1^-)})  \nonumber  \\
&+& e^{-i\widetilde{x}_L p^+(y^--y_1^-)}
(1-e^{-i\widetilde{x}_L p^+y_2^-})]  \nonumber  \\
&+&c_3(z,\ell_T^2,M^2)
e^{-i\widetilde{x}_L p^+(y^--y_1^-)}e^{-i\widetilde{x}_L xp^+y_2^-}
\label{hcd}
\end{eqnarray}
and
\begin{eqnarray}
c_1&=&
1+\frac{(1-z)^2(z^2-6z+1) }{1+z^2} \frac{M^2}{\ell_T^2} +
\frac{2z(1-z)^4}{1+z^2}\frac{M^4}{\ell_T^4}  \, , \label{c1}\\
 c_2&=&\frac{(1-z)}{2}
 \left\{1-\left[\frac{(1-z)(2z^3-5z+8z-1)}{(1+z^2)}\right.\right.
 \nonumber \\
 &+&\left.\frac{2C_F}{C_A}(1-z)^3\right]\frac{M^2}{\ell_T^2}
-\left[\frac{z(1-z)^4(3z-1)}{(1+z^2)}\right. \nonumber \\
&+& \left.\left.\frac{2C_F}{C_A}
 \frac{(1-z)^7}{(1+z^2)}\right]\frac{M^4}{\ell_T^4} \right\} \, ,
   \label{c2} \\
c_3&=&\frac{C_F (1-z)^2}{C_A}
\left[1-\frac{8z(1-z)^2}{1+z^2}\frac{M^2}{\ell_T^2} \right.\nonumber \\
& -& \left. \frac{(1-z)^4(z^2-4z+1)}{1+z^2}\frac{M^4}{\ell_T^4} \right]  
\, , \label{c3} 
\end{eqnarray}
where, $\widetilde{x}_L\equiv x_L+(1-z)x_M/z$ is the additional
fractional momentum of the initial quark or gluon in the 
rescattering that is required for gluon radiation,
and $x_L=\ell_T^2/2p^+q^-z(1-z)$.
The contribution from gluon fragmentation is similar to
that from quark fragmentation with $z\rightarrow 1-z$.
The virtual correction can be obtained via unitarity constraint.
One can recover the results for light quark rescattering \cite{ZW} \
by setting $M=0$ in the above equations. Notice that we have embedded the
phase factors from the LPM interference in the effective twist-four
parton matrix $T^{A,C}_{qg}(x,x_L,M^2)$.

Rewriting the sum of single and double scattering contributions
in a factorized form for the semi-inclusive hadronic tensor, one 
can define a modified effective fragmentation 
function $\widetilde{D}_{Q\rightarrow H}(z_H,\mu^2)$ as
\begin{eqnarray}
\widetilde{D}_{Q\rightarrow H}(z_H,\mu^2)&\equiv& D_{Q\rightarrow H}(z_H,\mu^2)
+\int_0^{\mu^2} \frac{d\ell_T^2}{\ell_T^2+(1-z)^2 M^2}  \nonumber \\
&&\hspace{-0.8in}\times\frac{\alpha_s}{2\pi}\int_{z_h}^1 \frac{dz}{z}
\Delta\gamma_{q\rightarrow qg}(z,x,x_L,\ell_T^2,M^2)
D_{Q\rightarrow H}(\frac{z_H}{z})  \nonumber \\
&&\hspace{-0.5in}+\int_0^{\mu^2} \frac{d\ell_T^2}{\ell_T^2+z^2 M^2}
\frac{\alpha_s}{2\pi}\int_{z_h}^1 \frac{dz}{z} \nonumber \\
&&\hspace{-0.5in}\times\Delta\gamma_{q\rightarrow gq}(z,x,x_L,\ell_T^2,M^2)
D_{g\rightarrow H}(\frac{z_H}{z}) \, , \label{eq:MDq}
\end{eqnarray}
where $D_{Q\rightarrow H}(z_H,\mu^2)$ and $D_{g\rightarrow
H}(z_H,\mu^2)$ are the leading-twist fragmentation functions of
the heavy quark and gluon. The modified splitting functions are given as
\begin{eqnarray}
\Delta\gamma_{q\rightarrow qg}(z)&=&
\left[\frac{1+z^2}{(1-z)_+}T^{A,C}_{qg}(x,x_L,M^2)\right. \nonumber \\
 &+&\left.\delta(1-z)\Delta
T^{A,C}_{qg}(x,\ell_T^2,M^2) \right] \nonumber \\
&&\hspace{-0.5in}\times \frac{2\pi C_A\alpha_s \ell_T^4}
{[\ell_T^2+(1-z)^2 M^2]^3 N_c f_q^A(x)} ,
\label{eq:r1}\\
\Delta T^{A,C}_{qg}(x,\ell_T^2,M^2) &\equiv & \int_0^1
\frac{dz}{1-z}\left[ 2 T^{A,C}_{qg}(x,x_L,m^2)|_{z=1}\right.
\nonumber \\
&& \hspace{-0.8in}\left. -(1+z^2)
T^{A,C}_{qg}(x,x_L,M^2)\right] \, , \label{eq:delta-T}
\end{eqnarray}
and $\Delta\gamma_{q\rightarrow gq}(z)=
\Delta\gamma_{q\rightarrow qg}(1-z)$. Here we have suppressed
other variables in $\Delta\gamma$.
Given the twist-four quark-gluon correction functions of the nucleus,
$T^{A,C}_{qg}(x,\ell_T^2,M^2)$, one should be able to evaluate the 
modified heavy-quark fragmentation function.

As seen from the phase factors in the
effective twist-four matrix element Eq.~(\ref{hcd}), the gluon
formation time for radiation from a heavy quark is
\begin{equation}
\tau_f\equiv\frac{1}{p^+\widetilde{x}_L}
=\frac{2z(1-z)q^-}{\ell_T^2+(1-z)^2M^2},
\end{equation}
which is shorter than that for gluon radiation from a light quark.
This should have significant consequences for the effective modified
quark fragmentation function and the heavy quark energy loss.

As discussed previously \cite{OW}, one can assume a 
factorized form of the twist-four parton matrix 
\begin{eqnarray}
 T^{A,C}_{qg}(x,x_L,M^2)
&\approx& \frac{\widetilde{C}}{x_A} f_q^A(x) \left \{
(1-e^{-\widetilde{x}_L^2/x_A^2}) \right.  \nonumber \\
&\times&\left[c_1(z,\ell_T^2,M^2)-c_2(z,\ell_T^2,M^2)\right] \nonumber \\
&+&\left. \frac{c_3(z,\ell_T^2,M^2)}{2} \right\}, \label{modT2}
\end{eqnarray}
in the limit $x_L\ll x_T \ll x$. 
Here $x_A\equiv 1/m_NR_A$
and $x_T\equiv \langle k_T^2\rangle/2p^+q^-z$ is the momentum
fraction associated with the initial intrinsic transverse momentum.
The coefficient $\widetilde{C}\equiv 2C x_Tf^N_g(x_T)$ 
should in principle depend on $Q^2$.
With this simplified form of twist-four matrix, one can then calculate
the heavy quark energy loss, defined as the fractional
energy carried by the radiated gluon, 
\begin{eqnarray}
\langle\Delta z_g^Q \rangle(x_B,Q^2) &=& \frac{\alpha_s}{2\pi}
\int_0^{Q^2} d\ell_T^2
\int_0^1 dz 
\frac{\Delta\gamma_{q\rightarrow gq}(z)}
{\ell_T^2 +z^2 M^2}z
 \nonumber \\
&&\hspace{-0.7in}
=\frac{\widetilde{C}C_A\alpha_s^2 x_B}{N_c Q^2 \, x_A}\int_0^1 dz
\frac{1+z^2}{z(1-z)} \int_{\widetilde{x}_M}^{\widetilde{x}_{\mu}}
d\widetilde{x}_L\frac{(\widetilde{x}_L-\widetilde{x}_M)^2}
{{\widetilde{x}_L}^4}
\nonumber \\
&&\hspace{-0.7in}\times
\left\{\frac{1}{2}c_3(z,\ell_T^2,M^2)
+\left(1-e^{-{\widetilde{x}_L}^2/x_A^2}\right) \right.  \nonumber \\
&&\hspace{-0.7in}\times\left. 
\left[c_1(z,\ell_T^2,M^2)
-c_2(z,\ell_T^2,M^2)\right]\right\}, \label{eq:loss1}
\end{eqnarray}
where $\widetilde{x}_M=(1-z)x_M/z$
and $\widetilde{x}_{\mu}=\mu^2/2p^+q^-z(1-z)+\widetilde{x}_M$. 
Note that $\widetilde{x}_L/x_A=L_A^-/\tau_f$ with $L_A^-=R_Am_N/p^+$
the nuclear size in the chosen frame. The LPM interference
is clearly contained in the second term of the integrand that 
has a suppression factor $1-e^{-\widetilde{x}_L^2/x_A^2}$. 
The first term that is proportional to $c_3(z,\ell_T^2,M^2)$ 
corresponds to a finite contribution in the factorization limit. 
We have neglected such a term in the study of light quark 
propagation since it is proportional to $R_A$, as compared to
the $R_A^2$ dependence from the first term due to LPM effect. 
We have to keep the first term for heavy quark propagation since the second
term will have a similar nuclear dependence when the mass dependence
of the gluon formation time is important.

\begin{figure}
\centerline{\psfig{file=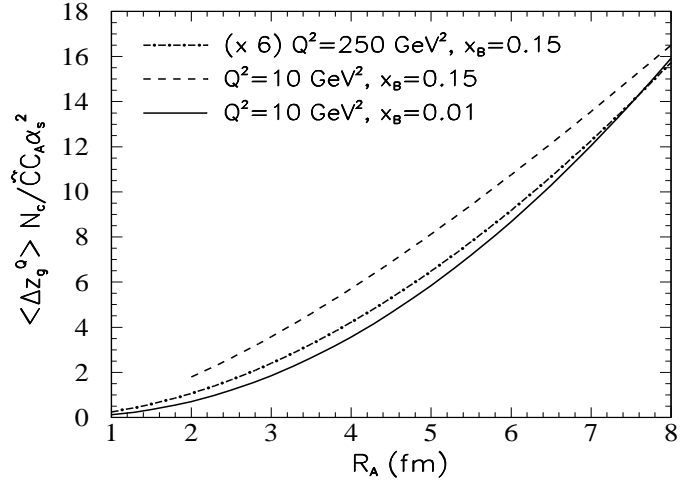,width=3.5in,height=2.5in}}
\caption{ The nuclear size, $R_A$, dependence of charm quark energy 
loss for different values of $Q^2$ and $x_B$.} 
\label{fig1}
\end{figure}

Since $\widetilde{x}_L/x_A\sim x_BM^2/x_AQ^2$, there are two
distinct limiting behaviors of the energy loss for different values
of $x_B/Q^2$ relative to $x_A/M^2$. When $x_B/Q^2\gg x_A/M^2$ for
small quark energy (large $x_B$) or small $Q^2$, the formation
time of gluon radiation off a heavy quark is always smaller than
the nuclear size. 
In this case, $1-\exp{(-{\widetilde{x}_L}^2/x_A^2)}\simeq 1$, so that 
there is no destructive LPM interference. The integral 
in Eq.~(\ref{eq:loss1}) is independent of $R_A$, and
the heavy quark energy loss 
\begin{equation}
\langle\Delta z_g^Q\rangle \sim C_A\frac{\widetilde{C}
\alpha_s^2}{N_c} \frac{x_B}{x_A Q^2}
\end{equation}
is linear in nuclear size $R_A$. In the opposite limit,
$x_B/Q^2\ll x_A/M^2$, for large quark energy (small $x_B$) or
large $Q^2$, the quark mass becomes negligible. The gluon formation
time could still be much larger than the nuclear size. The LPM
suppression factor $1-\exp{(-{\widetilde{x}_L}^2/x_A^2)}$ will
limit the available phase space for gluon radiation.
The integral in Eq.~(\ref{eq:loss1}) will be proportional
to $\int d\widetilde{x}_L[1-\exp(-{\widetilde{x}_L}^2/x_A^2)]
/{\widetilde{x}_L}^2 \sim 1/x_A$. 
The heavy quark energy loss
 \begin{equation}
\langle\Delta z_g^Q\rangle \sim C_A \frac{ \widetilde{C}
\alpha_s^2}{N_c} \frac{x_B}{x_A^2 Q^2}
\end{equation}
now has a quadratic dependence on the nuclear size similar to the
light quark energy loss.
Shown in Fig.~\ref{fig1} are the numerical results of the
$R_A$ dependence of charm quark energy loss, rescaled by
$\widetilde{C}(Q^2)C_A\alpha_s^2(Q^2)/N_C$, for different values
of $x_B$ and $Q^2$. One can clearly see that the $R_A$ dependence
is quadratic for large values of $Q^2$ or small $x_B$. The dependence
becomes almost linear for small $Q^2$ or large $x_B$. The charm quark
mass is set at $M=1.5$ GeV in the numerical calculation.

%%***********************************

\begin{figure}
\centerline{\psfig{file=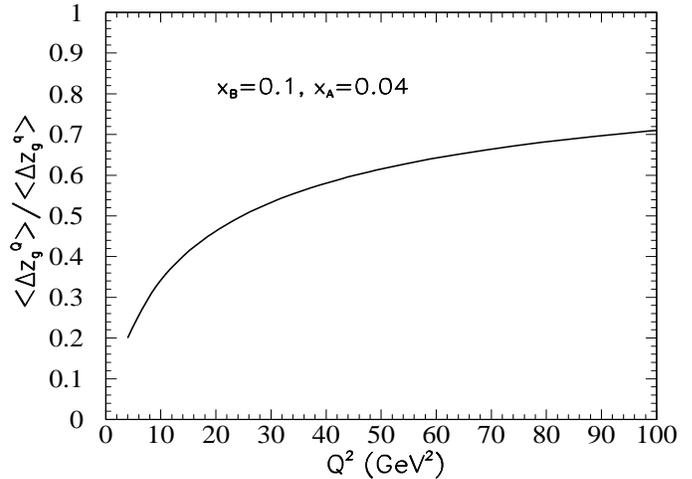,width=3.5in,height=2.5in}}
\caption{ The $Q^2$ dependence of the ratio between charm quark
and light quark energy loss in a large nucleus. } \label{fig2}
\end{figure}

\begin{figure}
\centerline{\psfig{file=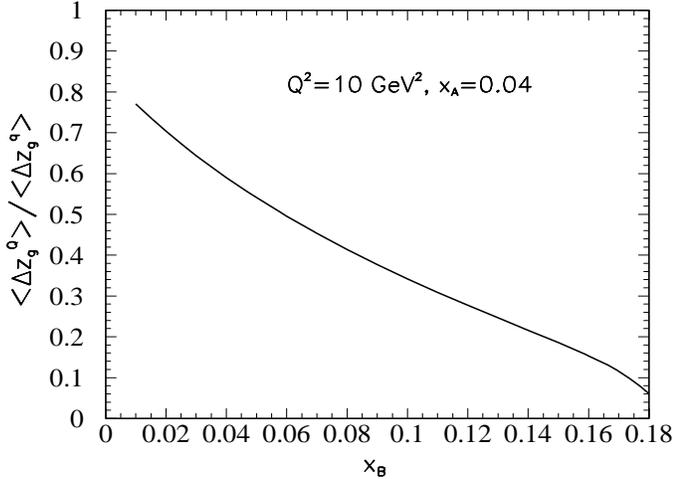,width=3.5in,height=2.5in}}
\caption{ The $x_B$ dependence of the ratio between charm quark
and light quark energy loss in a large nucleus. } \label{fig3}
\end{figure}

Another mass effect on the induced gluon radiation is the
``dead-cone'' phenomenon \cite{kharzeev} which suppresses the 
the small angle gluon radiation. Such a ``dead-cone'' effect is
manifested in Eq.~(\ref{wd1}) for the induced gluon spectra
from a heavy quark which is suppressed by a factor
\begin{equation}
f_{Q/q}=\left[\frac{\ell_T^2}{\ell_T^2+z^2
M^2}\right]^4=\left[1+\frac{\theta_0^2}{\theta^2}\right]^{-4},
\label{f}
\end{equation}
relative to that of a light quark for small angle radiation.
Here $\theta_0=M/q^-$ and $\theta=\ell_T/q^-z$.
This will lead to a reduced radiative energy loss of a heavy
quark, amid other mass dependence as contained in
$c_i(z,\ell_T^2,M^2)$ in Eqs.~(\ref{c1})-(\ref{c3}).
Setting $M=0$ in Eq.~(\ref{eq:loss1}), we recover the
energy loss for light quarks as in our previous study \cite{ZW}.
To illustrate the mass suppression of radiative energy loss
imposed by the ``dead-cone'', 
we plot the ratio 
$\langle\Delta z^Q_g\rangle(x_B,Q^2)/\langle\Delta z^q_g\rangle(x_B,Q^2)$
of charm quark and light quark energy loss as functions of $Q^2$ and $x_B$
in Figs.~\ref{fig2} and \ref{fig3}.

Apparently, the heavy quark energy loss induced by gluon radiation 
is significantly suppressed as compared to a light quark when 
the momentum scale $Q$ or the quark initial energy $q^-$ 
is not too large as compared to the quark mass. Only in the limit
$M \ll Q, \; q^-$, is the mass effect negligible. Then the energy loss
approaches that of a light quark.

In summary, we have calculated medium modification of fragmentation
and energy loss
of heavy quarks in DIS in the twist expansion approach. We demonstrated
that heavy quark mass not only suppresses small angle gluon radiation
due to the ``dead-cone'' effect but also reduces the gluon formation 
time. This leads to a reduced radiative energy loss as well as a 
different medium size dependence (close to linear), as compared to a 
light quark when the quark energy and the momentum scale $Q$ are of
the same order of magnitude as the quark mass. The result approaches
that for a light quark when the quark mass is negligible as compared to
the quark energy and the momentum scale $Q$. 
Similar to the case of light quark propagation, the result can 
be easily extended to a hot and dense medium, which will have
practical consequences for heavy quark production and suppression in
heavy ion collisions.

\section*{Acknowledgements}
This work was supported by the U.S.
Department of Energy under Contract No. DE-AC03-76SF00098
and by NSFC under project Nos. 19928511 and 10135030.
E.~Wang and B.~Zhang thank the Physics Department of Shandong
University for its hospitality during the completion of this work.

%%%%%%%%%%%%%%%%%%%%%%%%%%%%%%%%%%%%%%%%%%%

\end{multicols}


\begin{thebibliography}{99}

\bibitem{GW1}M. Gyulassy and X.-N. Wang, Nucl. Phys. {\bf B}420, 583 (1994)
[arXiv:nucl-th/9306003];
%%CITATION = NUCL-TH 9306003;%%
X.-N. Wang, M. Gyulassy and M. Pl\"umer, Phys. Rev. {\bf D} 51, 3436 (1995)
[arXiv:hep-ph/9408344].
%%CITATION = HEP-PH 9408344;%%

\bibitem{BDMPS} R. Baier {\it et al.}, Nucl. Phys. {\bf B483}, 291 (1997)
[arXiv:hep-ph/9607355];
%%CITATION = HEP-PH 9607355;%%
Nucl. Phys. {\bf B484}, 265 (1997)
[arXiv:hep-ph/9608322];
%%CITATION = HEP-PH 9608322;%%


\bibitem{Zh}B. G. Zhakharov, JETP letters {\bf 63}, 952 (1996)
[arXiv:hep-ph/9607440].
%%CITATION = HEP-PH 9607440;%%

\bibitem{GLV} M. Gyulassy, P. L\'evai and I. Vitev, Nucl. Phys.
{\bf B594}, 371 (2001)
[arXiv:nucl-th/0006010];
%%CITATION = NUCL-TH 0006010;%%
Phys. Rev. Lett. {\bf 85}, 5535 (2000)
[arXiv:nucl-th/0005032].
%%CITATION = NUCL-TH 0005032;%%

\bibitem{Wie}U. Wiedemann, Nucl. Phys. {\bf B588}, 303 (2000)
[arXiv:hep-ph/0005129];
%%CITATION = HEP-PH 0005129;%%
Nucl.\ Phys.\ A {\bf 690}, 731 (2001)
[arXiv:hep-ph/0008241].
%%CITATION = HEP-PH 0008241;%%


\bibitem{GuoW} X.~F.~Guo and X.-N.~Wang,
%``Multiple scattering, parton energy loss and modified
%fragmentation  functions in deeply inelastic e A scattering,''
Phys.\ Rev.\ Lett.\  {\bf 85}, 3591 (2000) [arXiv:hep-ph/0005044];
X.-N.~Wang and X.~F.~Guo,
%``Multiple parton scattering in nuclei: Parton energy loss,''
Nucl.\ Phys.\ A {\bf 696}, 788 (2001) [arXiv:hep-ph/0102230].

\bibitem{gvw}
M.~Gyulassy, I.~Vitev and X.~N.~Wang,
%``High p(T) azimuthal asymmetry in noncentral A + A at RHIC,''
Phys.\ Rev.\ Lett.\  {\bf 86}, 2537 (2001)
[arXiv:nucl-th/0012092].
%%CITATION = NUCL-TH 0012092;%%

\bibitem{ww02}
E.~Wang and X.~N.~Wang,
%``Jet tomography of dense and nuclear matter,''
Phys.\ Rev.\ Lett.\  {\bf 89}, 162301 (2002)
[arXiv:hep-ph/0202105].
%%CITATION = HEP-PH 0202105;%%

\bibitem{cw}C.~A.~Salgado and U.~A.~Wiedemann,
%``A dynamical scaling law for jet tomography,''
Phys.\ Rev.\ Lett.\  {\bf 89}, 092303 (2002)
[arXiv:hep-ph/0204221].
%%CITATION = HEP-PH 0204221;%%

\bibitem{phenix}
K.~Adcox {\it et al.}  [PHENIX Collaboration],
%``Suppression of hadrons with large transverse momentum in 
%central  Au + Au collisions at s**(1/2)(N N) = 130-GeV,''
Phys.\ Rev.\ Lett.\  {\bf 88}, 022301 (2002)
[arXiv:nucl-ex/0109003];
%%CITATION = NUCL-EX 0109003;%%
S.~S.~Adler {\it et al.}  [PHENIX Collaboration],
%``Suppressed pi0 production at large transverse momentum 
%in central  Au + Au collisions at s(NN)**(1/2) = 200-GeV,''
arXiv:nucl-ex/0304022.
%%CITATION = NUCL-EX 0304022;%%

\bibitem{star}
C.~Adler {\it et al.},
%``Centrality dependence of high p(T) hadron suppression in Au + Au  
%collisions at s(NN)**(1/2) = 130-GeV,''
Phys.\ Rev.\ Lett.\  {\bf 89}, 202301 (2002)
[arXiv:nucl-ex/0206011];
%%CITATION = NUCL-EX 0206011;%%
J.~Adams {\it et al.}  [STAR Collaboration],
%``Transverse momentum and collision energy dependence of high p(T) 
%hadron  suppression in Au + Au collisions at ultrarelativistic energies,''
arXiv:nucl-ex/0305015.
%%CITATION = NUCL-EX 0305015;%%

\bibitem{wang03}
X.~N.~Wang,
%``High p(T) hadron spectra, azimuthal anisotropy and back-to-back  
%correlations in high-energy heavy-ion collisions,''
arXiv:nucl-th/0305010.
%%CITATION = NUCL-TH 0305010;%%

\bibitem{kharzeev}
Y.~L.~Dokshitzer and D.~E.~Kharzeev,
%``Heavy quark colorimetry of QCD matter,''
Phys.\ Lett.\ B {\bf 519}, 199 (2001)
[arXiv:hep-ph/0106202].
%%CITATION = HEP-PH 0106202;%%

\bibitem{phenix2}
K.~Adcox {\it et al.}  [PHENIX Collaboration],
%``Measurement of single electrons and implications for charm 
%production  in Au + Au collisions at s(NN)**(1/2) = 130-GeV,''
Phys.\ Rev.\ Lett.\  {\bf 88}, 192303 (2002)
[arXiv:nucl-ex/0202002].
%%CITATION = NUCL-EX 0202002;%%

\bibitem{gyucharm} M.~Djordjevic and M.~Gyulassy,
%``Where is the charm quark energy loss at RHIC?,''
Phys.\ Lett.\ B {\bf 560}, 37 (2003)
[arXiv:nucl-th/0302069].
%%CITATION = NUCL-TH 0302069;%%

\bibitem{LQS}
 M. Luo, J. Qiu and G. Sterman, Phys. Lett. {\bf B279}, 377 (1992);
 M. Luo, J. Qiu and G. Sterman, Phys. Rev. D{\bf 50}, 1951 (1994);
 M. Luo, J. Qiu and G. Sterman, Phys. Rev. D{\bf 49}, 4493 (1994).

\bibitem{heavy2} B. Zhang, E. Wang and X. N. Wang, in preparation

\bibitem{ZW}
B.~W.~Zhang and X.~N.~Wang,
%``Multiple parton scattering in nuclei: Beyond helicity 
%amplitude  approximation,''
Nucl.\ Phys.\ A {\bf 720}, 429 (2003)
[arXiv:hep-ph/0301195].
%%CITATION = HEP-PH 0301195;%%


\bibitem{OW} J.~Osborne and X.-N.~Wang,
%``Multiple parton scattering in nuclei: Twist-four nuclear
%matrix  elements and off-forward parton distributions,''
Nucl.\ Phys.\ A {\bf 710}, 281 (2002)
[arXiv:hep-ph/0204046].
%%CITATION = HEP-PH 0204046;%%

\end{thebibliography}
\end{document}